\documentclass[%
 aip,
 apl,
 amsmath,amssymb,
 reprint,
]{revtex4-1}
\usepackage{graphicx}
\usepackage{dcolumn}
\usepackage{bm}
\usepackage[utf8]{inputenc}
\usepackage[T1]{fontenc}
\usepackage{booktabs} 
\usepackage{mathptmx}
\usepackage{etoolbox}
\usepackage{hyperref}
\usepackage{xcolor}
\usepackage{braket}

\usepackage{subcaption}

\makeatletter
\def\@email#1#2{%
 \endgroup
 \patchcmd{\titleblock@produce}
  {\frontmatter@RRAPformat}
  {\frontmatter@RRAPformat{\produce@RRAP{*#1\href{mailto:#2}{#2}}}\frontmatter@RRAPformat}
  {}{}
}%
\makeatother

\begin{document}

\title{Nitrogen Vacancy Centers in Hexagonal Diamond Exhibit Long Coherence Times}

\author{Gabriel Kumar}
\affiliation{Pritzker School of Molecular Engineering, The University of Chicago, Chicago, Illinois 60637, United States}

\author{Siyuan Chen}
\affiliation{Pritzker School of Molecular Engineering, The University of Chicago, Chicago, Illinois 60637, United States}

\author{Victor Wen-zhe Yu}
\affiliation{Materials Science Division, Argonne National Laboratory, Lemont, Illinois 60439, United States}

\author{Giulia Galli}
\email[]{gagalli@uchicago.edu} 
\affiliation{Pritzker School of Molecular Engineering, The University of Chicago, Chicago, Illinois 60637, United States}
\affiliation{Materials Science Division, Argonne National Laboratory, Lemont, Illinois 60439, United States}
\affiliation{Department of Chemistry, The University of Chicago, Chicago, Illinois 60637, United States}
\date{\today}

\begin{abstract}
We show that negatively charged nitrogen-vacancy (NV) centers in the hexagonal diamond polymorph lonsdaleite offer a route to spin qubits with enhanced coherence relative to their cubic-diamond counterparts. Using first-principles calculations, we examine two distinct defect configurations, AA, with the same symmetry as in cubic diamond and AB, with reduced symmetry. We find that the AB configuration of the NV center exhibits a finite transverse zero-field splitting, giving rise to an approximate fourfold enhancement of the Hahn-echo coherence time $T_2$ at zero magnetic field. The AA configuration, by contrast, closely reproduces the electronic structure and coherence properties of the cubic NV center. We further characterize the many-body electronic structure, vertical excitation energies, and photoluminescence spectra of both configurations, providing spectral fingerprints for their experimental identification. Our results establish symmetry-broken NV centers in lonsdaleite as promising candidates for quantum sensing and information science applications.
\end{abstract}

\maketitle

Negatively charged nitrogen-vacancy (NV) centers in cubic diamond are prototypical spin qubits, with applications in quantum technologies, including sensing\cite{Abobeih2019-db, Du2024-be}, computing\cite{Bartling2025-ph, Bradley2019-ae}, and communication\cite{Kimble2008-hg, Wehner2018-lm}. They can be optically initialized, with the excitation from the ground to the first excited triplet state decaying non-radiatively to singlet shelving states and eventually to the ground state with a specific magnetic number $m_s$. The qubit can then be interrogated with microwave radiation\cite{Robledo2011-yi}.

To date, most studies have focused on NV centers in the cubic diamond lattice. However, several recent theoretical studies have investigated NV centers in hexagonal diamond (also called lonsdaleite). This polymorph was first observed in 1967 inside the Canyon Diablo meteorite\cite{Frondel1967-mi}; however its synthesis in the laboratory has been achieved only recently\cite{Lai2026-bk, Yang2025-uc}. 
Moreover, there has been a recent synthesis of NV centers in rhombohedral diamond under conditions similar to those used to synthesize lonsdaleite\cite{Dalmieda2026-kq}. 

\begin{figure*}[ht!]

    \begin{subfigure}{0.18\textwidth}
        \includegraphics[width=\textwidth]{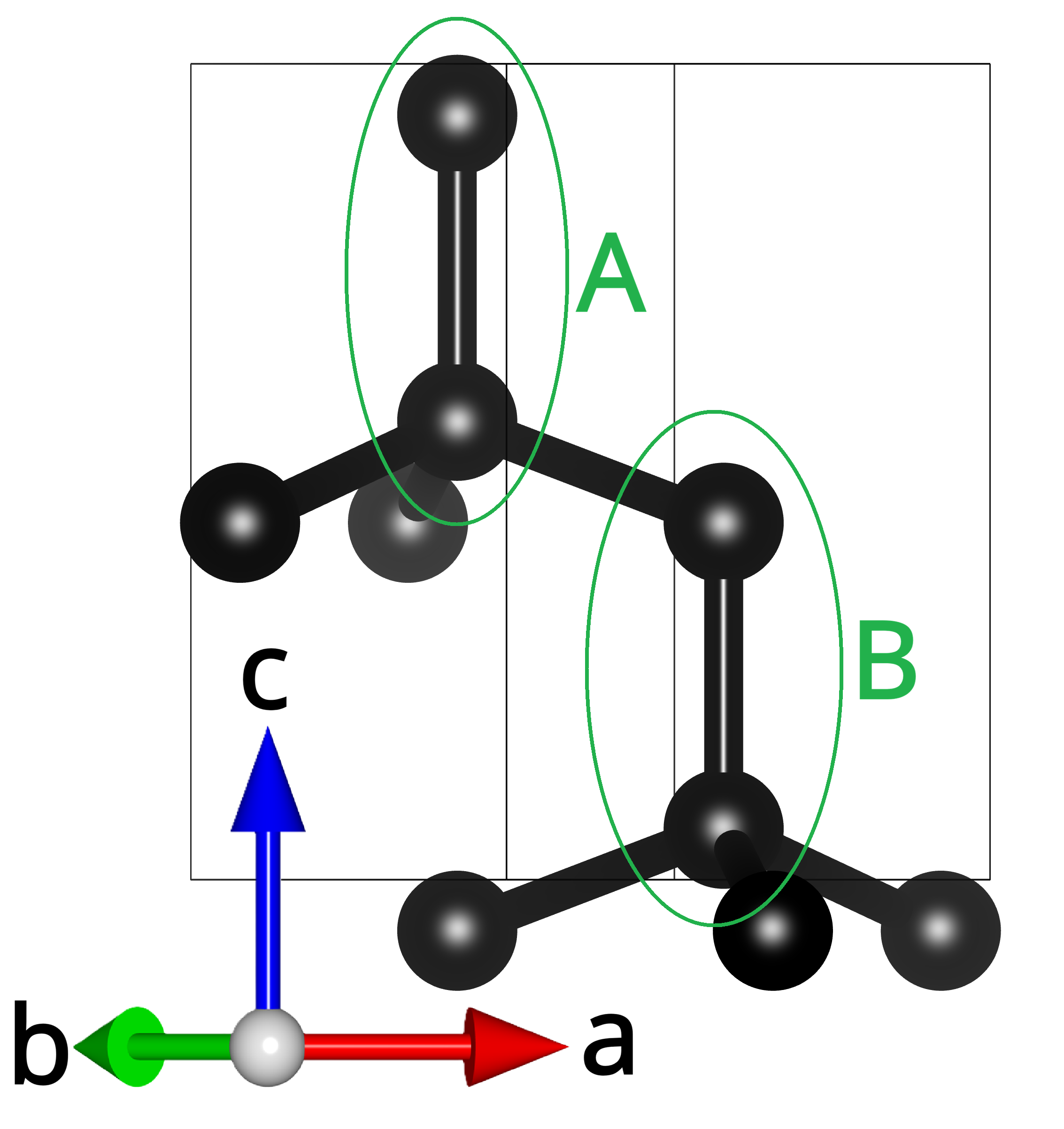}
    \end{subfigure}
    \hspace{15pt}
    \begin{subfigure}{0.3\textwidth}
        \includegraphics[width=\textwidth]{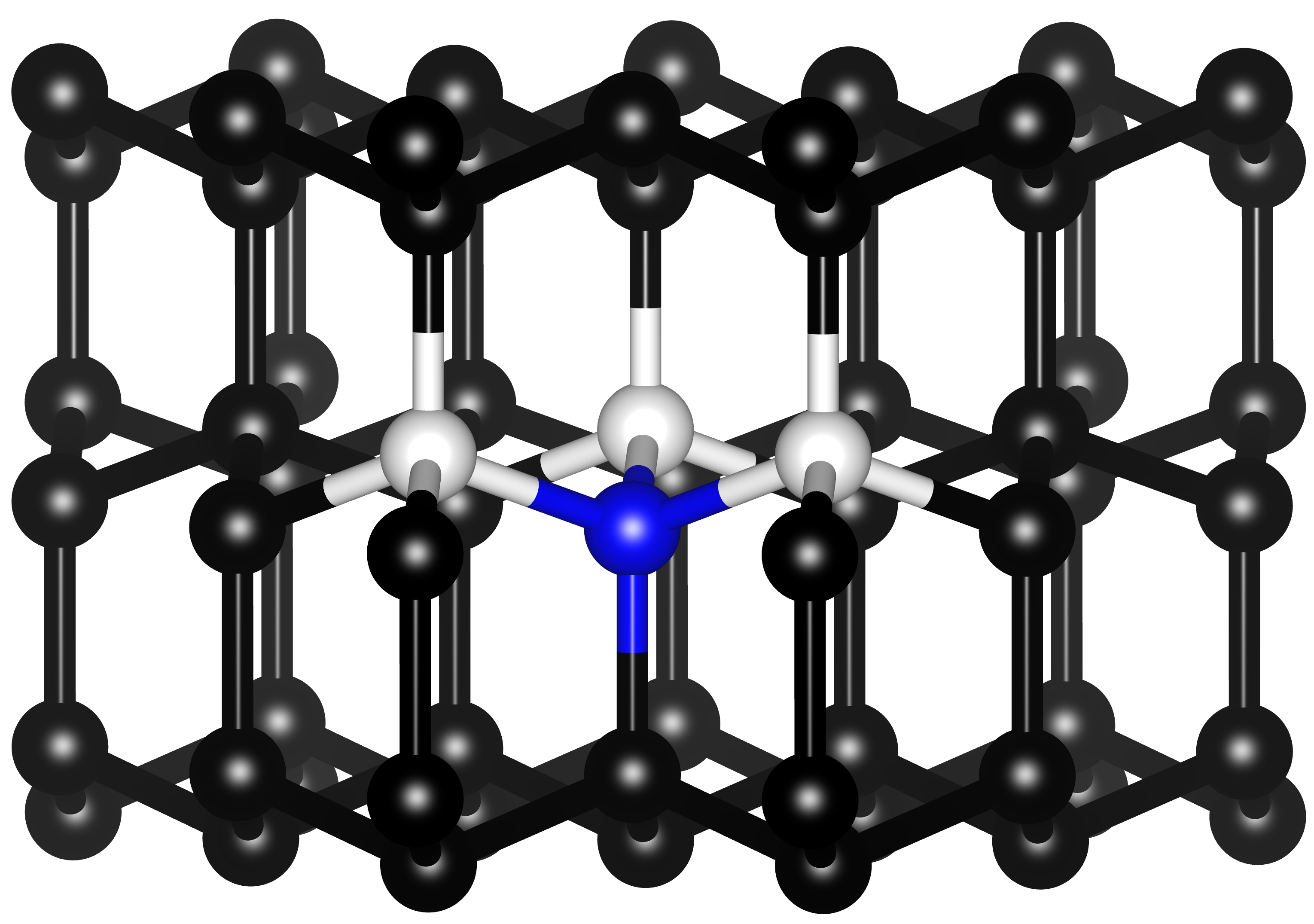}
    \end{subfigure}
    \hspace{15pt}
    \begin{subfigure}{0.24\textwidth}
        \includegraphics[width=\textwidth]{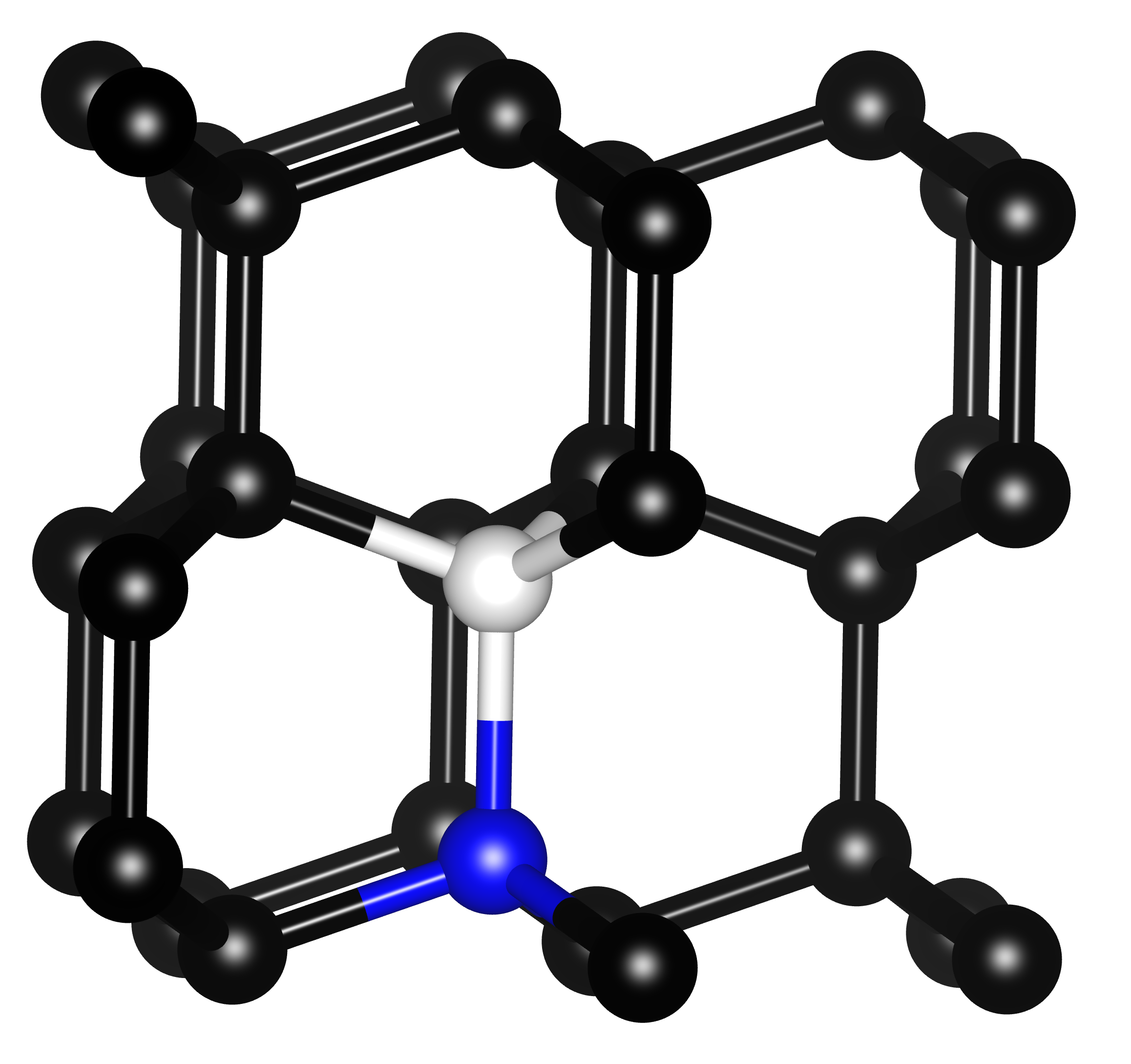}
    \end{subfigure}

    \caption{Crystal structure of hexagonal diamond (lonsdaleite), with point group P6$_3$/mmc. (Left) The 4-atom unit cell, with experimental lattice parameters $a$ = 2.51 \AA\ and $c$ = 4.12 \AA, compared to 
    $a$ = 2.517 \AA\ and $c$ = 6.166 \AA\ for cubic diamond when represented in a hexagonal lattice. The two pairs of carbon atoms aligned along the $c$-axis are labeled A and B. (Middle) The AB configuration of the NV center with the NV axis aligned along the bond connecting A and B. The three symmetry-equivalent orientations are shown. (Right) The AA configuration of the NV center with the NV axis aligned along a bond within A or B. Carbon, nitrogen, and vacancy sites are shown as black, blue, and white spheres, respectively.}
    \label{fig:geometry}
\end{figure*}

A ball-and-stick representation of the possible NV configurations in lonsdaleite is given in Fig. \ref{fig:geometry}, where they are labeled AB and AA, according to the relative positions of the nitrogen atom and vacancy within the host lattice. The AA configuration preserves the $C_{3v}$ symmetry of the NV center in cubic diamond (referred to as a cubic NV hereafter), while the AB configuration exhibits a lower $C_s$ symmetry. 

Using density functional theory (DFT) calculations, Sun et al. \cite{Sun2024-in} investigated the band structure and phonon spectra of NV centers near the surface of lonsdaleite, and concluded that depending on the surface termination, NV centers may retain the desired negative charge and spin-triplet state. Subsequently, Abdelghafar et al. \cite{abdelghafar_density_2025,Abdelghafar2025-pm} investigated the photoluminescence (PL) spectra of these defects in bulk lonsdaleite, showing differences for centers in the AB and AA configurations due to changes in the phonon spectrum. Manian et al. \cite{Manian2025-va} also performed DFT calculations on nanoscale lonsdaleite to identify the spectral fingerprints of the NVs in the various configurations. 
However, the excited-state properties, many-body electronic structure, and spin coherence of NVs in lonsdaleite remain largely unexplored.

Here, using first-principles computations, we predict the ground- and excited-state electronic properties of NVs in lonsdaleite and we investigate their coherence properties. We use DFT for ground-state geometries, quantum defect embedding theory (QDET) to describe many-body electronic states\cite{Chen2025-cy, Sheng2022-yz}, the generating function approach for PL spectra\cite{PhysRevMaterials.5.084603}, and the generalized cluster-correlation expansion (gCCE) method for spin coherence times\cite{Onizhuk2025-cd}. As expected, we find that the properties of NV centers in the AA configuration closely resemble those of cubic NVs. However, NVs in the AB configuration with reduced symmetry exhibit a finite $E$ component of the zero-field splitting (ZFS) tensor, leading to spin coherence times at zero magnetic field approximately four times larger than those of cubic NVs. Interestingly, despite the symmetry breaking, the electronic structure of the AB center remains similar to that of the cubic NV. Our findings suggest that the AB NV centers in lonsdaleite are promising candidates for quantum information science applications.

We studied two supercells of lonsdaleite, with 299 and 575 atoms, respectively. We carried out DFT calculations with the Quantum ESPRESSO code\cite{Giannozzi2009-il, Giannozzi2017-kz}, using a 60 Ry kinetic energy cutoff for the plane-wave basis set, the Perdew-Burke-Ernzerhof (PBE) exchange-correlation functional\cite{Perdew1996-gi}, and the SG15 optimized norm-conserving Vanderbilt pseudo-potentials\cite{Hamann2013-py, Schlipf2015-xh}. The Brillouin zone of the supercells was sampled with the $\Gamma$ point. The atomic positions were relaxed within a fixed cell size. We found that the C-C first-neighbor distances in lonsdaleite differ by 0.01--0.02 \AA\ relative to those in cubic diamond, consistent with what is reported experimentally\cite{Yang2025-uc}.

Based on the defect formation energy calculations of Ref.~\onlinecite{Manian2025-va}, AA and AB NV centers in lonsdaleite can be stabilized in the experimentally relevant negatively charged ($-1$) spin-triplet state. 
We first discuss vertical excitation energies (VEEs) computed using QDET\cite{Chen2025-cy, Sheng2022-yz} as implemented in the WEST code\cite{Yu2022-cb, Yu2026-xr, Govoni2015-ve}. In QDET, 
an effective many-body Hamiltonian is derived for an active space consisting of a chosen subset of DFT orbitals describing the defect states. As in cubic diamond, the defect states of the NV in lonsdaleite are well localized within the host gap, which is found to be $\simeq$ 1 eV smaller than that of the cubic polymorph, in agreement with experiments and previous calculations\cite{Gao2015-re, Zheng2024-fi, Yelisseyev2020-ty}. 
Following Ref. \onlinecite{Chen2025-cy}, we considered active spaces containing all localized defect orbitals (minimum model), and occupied bands close in energy to the valence band maximum (VBM), specifically 3 eV below the VBM for the 575-atom cell. 
Diagonalization of the QDET effective Hamiltonian 
was carried out using the full configuration interaction (FCI) module of the PySCF code\cite{Sun2018-zu, Sun2020-hl}.

\begin{figure}[ht!]
    \includegraphics[width=0.49\textwidth]{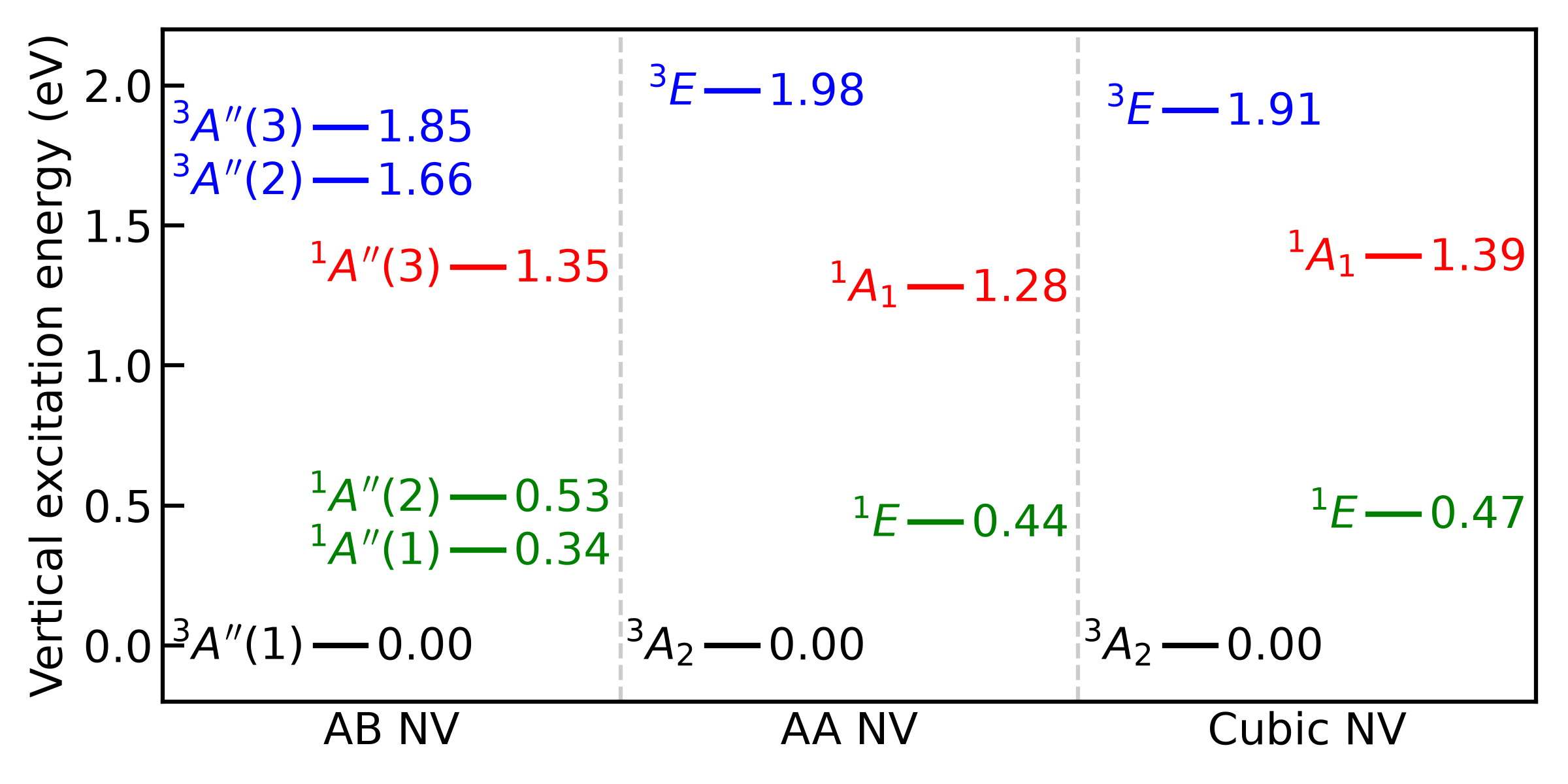}
    \caption{Vertical excitation energies of NV centers in lonsdaleite and cubic diamond computed with the quantum defect embedding theory (QDET) (see text). The many-body states are labeled according to the irreducible representations of the defect point groups ($C_s$ and $C_{3v}$ for AB and AA NV, respectively). States in the AA and cubic NVs with the same color have the same symmetry character. For the AB NV, the colors are the same as those of the corresponding states in the AA NV from which they split as the symmetry is reduced.}
    \label{fig:opticalCycle}
\end{figure}

The many-body states obtained from QDET calculations are shown in Fig. \ref{fig:opticalCycle}, and are labeled according to the irreducible representations of the NV-center point group. Calculations were conducted starting from Kohn-Sham wavefunctions at the PBE level of theory. It has been shown that starting from wavefunctions computed with hybrid functionals, e.g. the dielectric-dependent hybrid (DDH) functional~\cite{ddh_skone_2014,ddh_skone_2016} as in Ref.~\onlinecite{Chen2025-cy}, yields results in better agreement with experiments for the cubic NV center. However, here we opted for the PBE functional, as our focus is on the comparison of results between varied NV configurations, not on absolute numbers.

\par The NV center in cubic diamond exhibits a $^3A_2$ ground state and a $^3E$ excited state, with the singlet states $^1E$ and $^1A_1$ lying between the triplet manifolds. The AA configuration, which preserves $C_{3v}$ symmetry, displays a similar many-body electronic structure, with energy differences smaller than 0.1 eV. In contrast, the AB configuration has reduced $C_s$ symmetry, which lifts the degeneracies of the $^1E$ and $^3E$ states and results in splittings of 0.19 eV in both manifolds. The VEE from the $^3A''(1)$ ground state to the $^3A''(2)$ excited state is 1.66 eV, about 0.3 eV smaller than the corresponding VEE (1.98 eV) of the lowest triplet excited state in the AA configuration.

Given the resemblance between the many-body electronic-state diagrams of NVs in cubic diamond and lonsdaleite, we expect the latter to support an optical cycle similar to that of the cubic polymorph. 
Nevertheless, establishing the efficiency of such cycle would require a calculation of intersystem crossing rates~\cite{Jin2025-kl, Zhang2026-wi, Huang:2025vzx}, which are beyond the scope of the present work. However, we note that symmetry-breaking configurations similar to the AB NV have also been reported for NV centers in cubic diamond in the vicinity of dislocations.
For a representative set of those configurations, Zhang et al.~\cite{Zhang2026-wi} computed the intersystem crossing rates and simulated optical spin initialization and readout processes, confirming the feasibility of an optical cycle under symmetry-lowered conditions.

We now turn to describe the PL spectra, computed using the PyPL code~\cite{jin2022vibrationally, PhysRevMaterials.5.084603} and employing the 575-atom supercell. Vibronic effects were treated within the one-dimensional configuration-coordinate approximation (1D CCD), in which the multidimensional phonon problem is reduced to the calculation of a single effective mode defined by the linear interpolation between ground- and excited-state equilibrium geometries; these were obtained using DFT and the $\Delta$SCF approach, respectively.
The PL spectra are evaluated using Fermi's golden rule and a generating function approach within the Huang-Rhys (HR) formalism\cite{Huang1950-sh,PhysRevMaterials.5.084603}. 

\begin{figure}[ht!] 
    \centering

    \begin{subfigure}{\linewidth}
        \centering
        \includegraphics[width=0.85\linewidth]{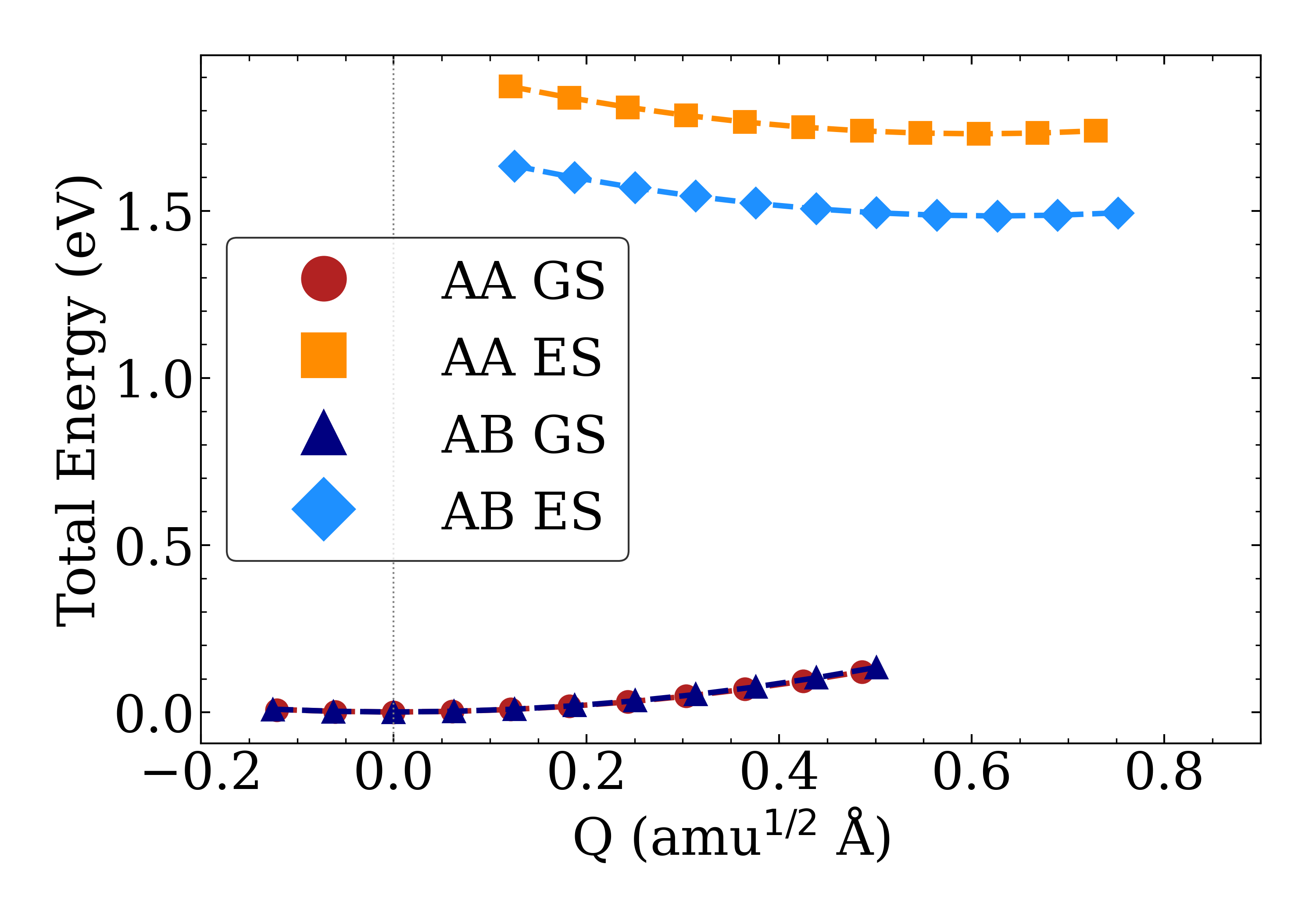}
        \caption{}
        \label{fig:spectrumAA}
    \end{subfigure}

    \vspace{0.1cm} 

    \begin{subfigure}{\linewidth}
        \centering
        \includegraphics[width=0.85\linewidth]{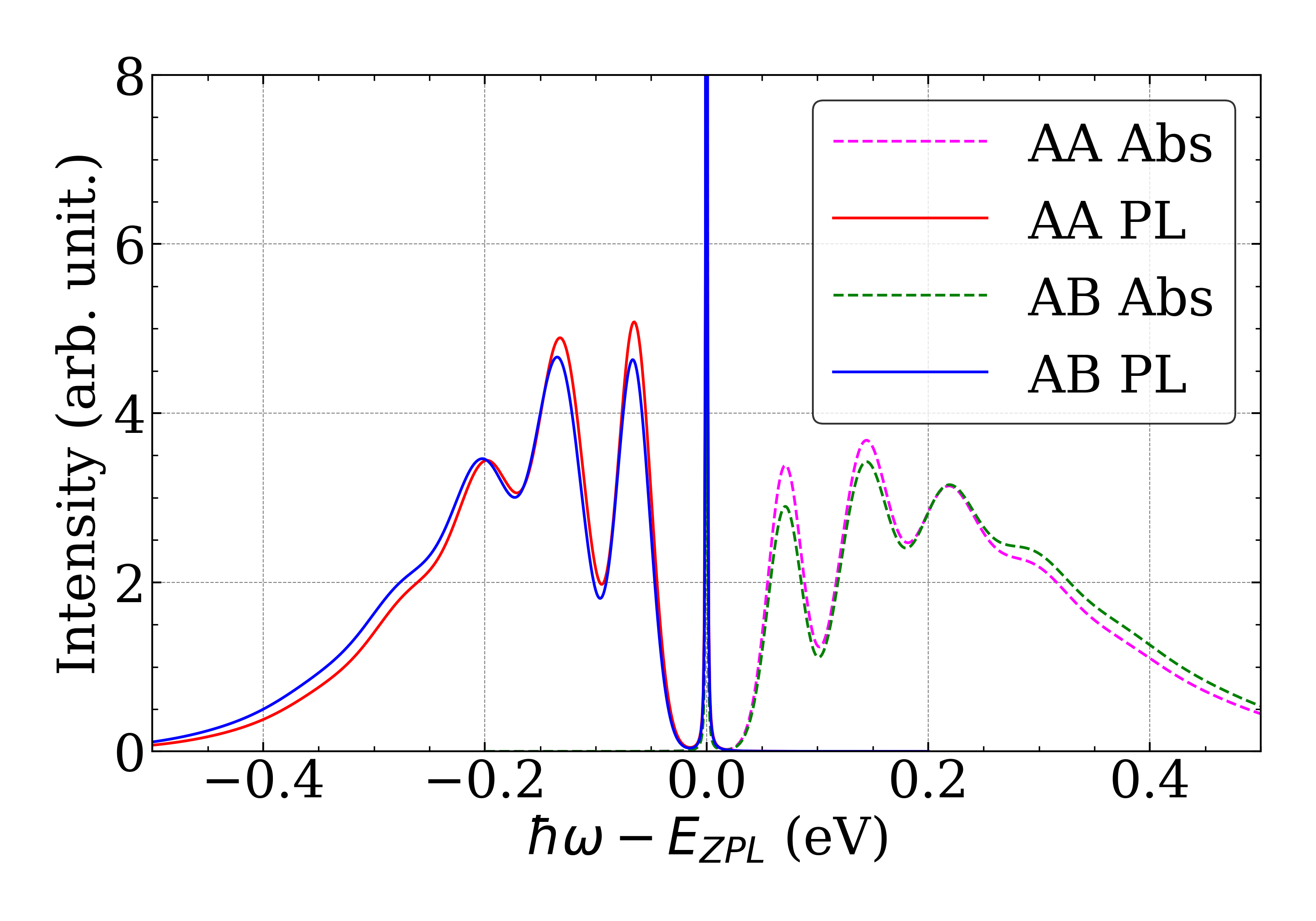}
        \caption{}
        \label{fig:spectrumAB}
    \end{subfigure}

    \caption{(a) Configuration-coordinate diagrams obtained from $\Delta$SCF calculations for the NV AA and AB configurations, showing the interpolation between the ground- (zero of energy) and excited-state geometries. (b) Simulated photoluminescence (PL) and absorption (Abs) spectra at $T=5$ K, computed using the zero-phonon line and phonon modes extracted from the configuration-coordinate diagrams. The x-axis shows the energy shift relative to the ZPL energies of 1.48 and 1.73 eV for the AB and AA configurations, respectively. The fitting parameters are the same as those used for the NV center in cubic diamond~\cite{PhysRevMaterials.5.084603}.}
    \label{fig:plSpectrum}
\end{figure}

The energies of the ground- and excited-state configurations used to construct the effective phonon modes are shown in Fig.~\ref{fig:plSpectrum}. From these modes, we computed the HR and Debye-Waller (DW) factors, reported in Table~\ref{tab:nv_comparison}.
Both the AA and AB configurations exhibit larger DW factors (4.42\% and 3.65\%, respectively) than the cubic NV (2.42\%), indicating a stronger zero-phonon line (ZPL) contribution than in the cubic case. Our computed DW values are in good agreement with those of Ref.~\onlinecite{Abdelghafar2025-pm}: 4.5--5.4\% for the AA configuration and 2.3--2.5\% for the AB configuration. 

\begin{table*}[!ht]
\centering
\caption{Comparison of computed optical and spin properties for AB, AA, and cubic NV centers, together with values from the literature. The AB $T_2$ is the average of the linear and quadratic interpolation schemes (see text). The experimental values are for cubic NVs. The agreement with experiment could be improved (see text) using hybrid functionals instead of PBE, chosen here for simplicity as we focus on comparisons between hexagonal and cubic diamond.}
\begin{tabular}{|l|ccc|c|}
\hline
\textbf{Property} & \textbf{AB} & \textbf{AA} & \textbf{Cubic} & \textbf{Experiment} \\
\hline
First Triplet Excitation [eV] & 1.66 & 1.98 & 1.90\cite{Chen2025-cy} & 2.18\cite{Davies1976-ez}\\
Zero-Phonon Line [eV] & 1.48 & 1.73 & 1.73 \cite{Jin2023-tr} & 1.945\cite{Doherty2013-ms} \\
Huang-Rhys Factor (Excited State) & 3.31 & 3.12 & 3.72 & 3.87\cite{Lamelas2024-qo} \\
Debye-Waller Factor (Excited State) & 3.65\% & 4.42\% & 2.42\% & 2.09\%\cite{Lamelas2024-qo} \\
Zero-Field Splitting $D$ [GHz] & 2.85 & 2.84 & 3.04 & 2.87\cite{Fuchs2008-wk} \\
Zero-Field Splitting $E$ [MHz] & -358.27 & -3.05 & -0.15 & 0 \\
Coherence Time ($T_2$) [ms] & 3.73 & 0.9 & 0.85 & 0.1-1.8\cite{Maze2008-ts, Mizuochi2009-sd, Balasubramanian2009-qq}\\
\hline
\end{tabular}
\label{tab:nv_comparison}
\end{table*}

The computed PL spectra are reported in Fig.~\ref{fig:plSpectrum}, showing that the AB and AA configurations have similar line shapes, albeit with different ZPL energies (see Table~\ref{tab:nv_comparison}). Note that the PBE functional used here is known to underestimate the ZPL energy of NVs in cubic diamond, compared, for example, to the hybrid functional DDH. However, as mentioned above, our main interest is to investigate trends and differences between AA, AB, and cubic NVs; based on our previous results~\cite{Jin2023-tr}, trends obtained with PBE are expected to be as accurate as those of DDH. Using $\Delta$SCF-PBE, we predict the ZPL energy of the AB configuration to be 0.25 eV lower than that of the AA configuration, consistent with the difference in VEEs obtained with QDET. This trend is also consistent with the results of Ref.~\onlinecite{Abdelghafar2025-pm}, who performed $\Delta$SCF calculations using the HSE06 hybrid functional. 
However, our results differ from those reported in Ref.~\onlinecite{Manian2025-va}, which employs the vertical gradient approximation within DFT,
the PBE0 hybrid functional with a Becke-Johnson dispersion correction and a 220-atom nanocrystal model. The authors reported the AA NV having a lower ZPL energy than the AB NV. In addition to a different functional, this difference in ordering may be due to the use of NV in nanoclusters, with specific surface terminations and vibrational properties different from those of bulk diamond.

We note that the ZPL of the AA configuration obtained with $\Delta$SCF-PBE is identical to that predicted for the cubic NV using the same method \cite{Jin2023-tr}, which is known to underestimate experiments by approximately 0.2 eV. Hence we expect our $\Delta$SCF-PBE calculations to do the same for the AA and AB NVs.

We conclude our analysis of the NV properties by discussing coherence times. We first computed the ZFS tensor using the PyZFS code~\cite{Ma2020} with ground-state wavefunctions obtained from a 575-atom supercell, without including spin–orbit coupling effects, which were reported to be negligible for NV in cubic diamond\cite{Biktagirov2020-sc}. Coherence functions were then computed using the second-order gCCE method, as implemented in the PyCCE code~\cite{https://doi.org/10.1002/adts.202100254}. Based on convergence tests, dipolar interaction cutoffs of 10 and 8 \AA\ and nuclear-spin bath spins within outer radii of 50 and 40 \AA\ were included for the AA and AB configurations, respectively. We also performed simulations on a 511-atom cubic NV cell with the same parameters as the AA NV. 
At zero magnetic field ($B$), for the AB configuration, we estimated the coherence function using an average of linear and quadratic interpolations. 

\begin{figure}[ht!]
    \includegraphics[width=0.49\textwidth]{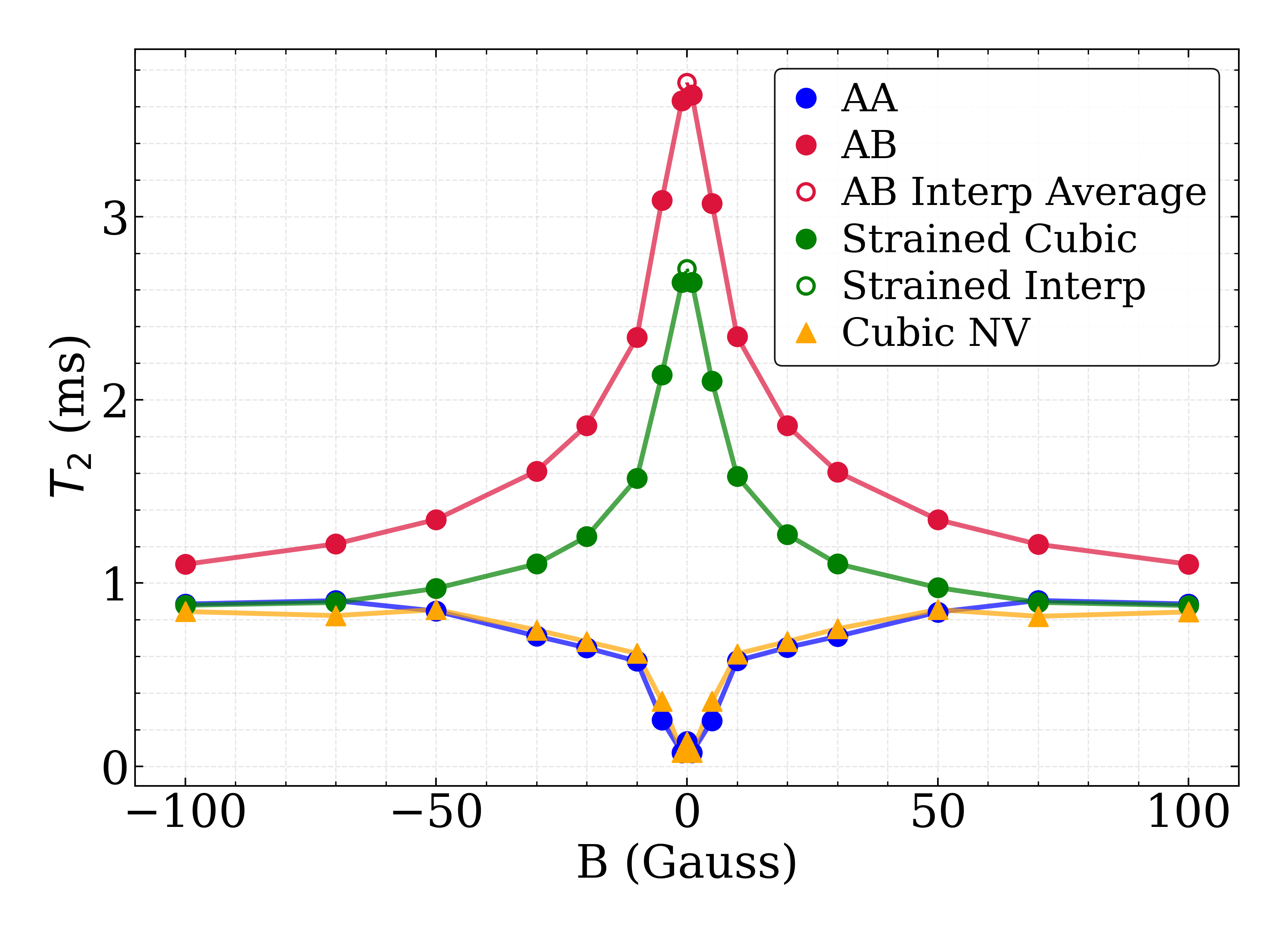}
    \caption{Computed coherence time $T_2$ as a function of the applied magnetic field for the AA and AB configurations in hexagonal diamond (lonsdaleite), the NV center in cubic diamond, and the NV center in cubic diamond with a $3\%$ tensile strain applied along an N-C bond. For the AB configuration and the strained cubic diamond NV, the $B=0$ data point is estimated by averaging the left- and right-biased linear and quadratic interpolations.}
    \label{fig:coherence}
\end{figure}

The computed coherence time $T_2$ as a function of the applied magnetic field is shown in Fig.~\ref{fig:coherence}. The AB configuration reaches a maximum $T_2$ of approximately 3.73 ms, whereas the AA configuration has a max $T_2$ of 0.9 ms at $B=70$ G that decreases at low magnetic fields similar to that of cubic NV, whose max $T_2$ = 0.85 ms at $B=50$ G. This pronounced difference between AA and AB arises from the presence of a nonzero $E$ component of the ZFS tensor in the AB configuration, which originates from the symmetry breaking discussed earlier. Specifically, the AB configuration has ZFS parameters $D=2.85$ GHz and $E=-358.27$ MHz, while the AA configuration has $D=2.84$ GHz and $E=-3.05$ MHz (with $E \approx 0$ enforced in our gCCE calculations for numerical stability). The cubic configuration has $D=3.04$ GHz and $E=-0.15$ MHz (with $E\approx 0$ enforced again). The difference in cubic and AA $E$ values is due to the difference in supercell geometry and a minor symmetry breaking. Our calculated $D$ values differ from those reported in Ref.~\onlinecite{Manian2025-va} ($D=2.74$ and $D=4.56$ GHz for the AB and AA configurations, respectively), where the ZFS was estimated from the change in the N-V bond length relative to the cubic NV using the strain susceptibility of the ground-state fine structure of cubic NV.
Since our calculations explicitly evaluate the ZFS from the electronic structure of each relaxed defect configuration, they are expected to provide a more accurate description of the ZFS.

The predicted enhancement of $T_2$ is consistent with the trend reported in Ref.~\onlinecite{Zhang2026-wi}, where symmetry breaking induced by dislocations in cubic diamond introduces a finite $E$ term in the ZFS tensor, leading to longer $T_2$ at low $B$. We recall that the enhancement can be easily understood by considering a spin Hamiltonian including the ZFS and Zeeman terms, and a magnetic field $\mathbf{B} = (0, 0, B_z)$ aligned with the NV quantization axis: 

\begin{equation}
H = DS_z^2 + E(S_x^2 - S_y^2) + \gamma_e B_z S_z \,,
\end{equation}
where $\gamma_e$ is the electron gyromagnetic ratio. When the transverse ZFS component $E \ne 0$, the degeneracy of the $\ket{\pm 1}$ states is lifted, and the eigenstates are given by the symmetric and antisymmetric superpositions $\ket{\pm} = \frac{1}{\sqrt{2}}\bigl(\ket{-1} \pm \ket{+1}\bigr)$, with transition frequencies between $\ket{\pm}$ and $\ket{0}$:
\begin{equation}
\omega_{0, \pm} = D \pm \sqrt{(\gamma_e B_z)^2 + E^2} \,.
\end{equation}
Hence the first derivative of $\omega_{0, \pm}$ with respect to $B_z$;
\begin{equation}
\frac{\partial \omega_{0, \pm}}{\partial B_z} = \pm \frac{\gamma_e^2 B_z}{\sqrt{(\gamma_e B_z)^2 + E^2}} \,,
\end{equation}
vanishes at $B_z = 0$, with 
the transition frequency being insensitive to magnetic noise to first order (clock transition). The sensitivity is determined by the second derivative
\begin{equation}
\frac{\partial^2 \omega_{0, \pm}}{\partial B_z^2} = \pm \frac{\gamma_e^2 E^2}{((\gamma_e B_z)^2 + E^2)^{3/2}} \,,
\end{equation}
which is $\pm \frac{\gamma_e^2}{E}$ at $B_z = 0$. Thus, increasing the $E$ component of the ZFS tensor 
reduces its sensitivity to second-order fluctuations in the magnetic field, leading to an extended spin coherence time.

\par A finite $E$ splitting can also be induced in cubic diamond through strain. As a representative case, we investigated a 511-atom supercell containing an NV center under a relatively large 3\% uniaxial tensile strain applied along an N-C bond direction, and found $E=-155.80$ MHz. This result compares favorably with that obtained with the strain-spin relationship derived by Udvarhelyi et al.\cite{Udvarhelyi2018-of} ($E=-130.68$ MHz). Hence, a value of $E$ roughly half of that of the AB configuration can be obtained in cubic diamond, although it requires large strains applied along specific crystallographic directions. The calculated $T_2$ for this strained sample is $2.71$ ms, smaller than that of hexagonal diamond. 

In summary, we investigated the properties of NV$^-$ centers in lonsdaleite using first-principles calculations, and reported the many-body electronic structure and spin coherence times of NVs in the AA and AB configurations. From gCCE calculations, we find that the AB NV center exhibits a peak coherence time of 3.73 ms, while the AA NV center reaches 0.9 ms, compared to 0.85 ms for the cubic NV$^-$ center. The enhanced coherence in the AB configuration is driven by a sizable $E$ component of the ZFS tensor arising from symmetry breaking.

Our calculations show that the electronic structure of the AB NV exhibits a splitting of the $^1E$ and $^3E$ states due to reduced symmetry, whereas that of the AA NV closely resembles the electronic structure of the cubic NV. We also computed the PL spectra, HR and DW factors, ZPL energies, and ZFS parameters for both configurations (Table~\ref{tab:nv_comparison}), which may aid their experimental identification. 

Our results indicate that NV centers in lonsdaleite are promising candidates for quantum applications. In particular, the AB configuration combines extended coherence times with electronic structure similar to that of the cubic NV. The coexistence of AA and AB centers within the same material may open the possibility 
of exploring NV ensembles with distinct but complementary properties not directly accessible in cubic diamond.

\begin{acknowledgments}
We thank Jonah Nagura and Michael Toriyama for useful discussions. The computational part of our research was supported by the Midwest Integrated Center for Computational Materials (MICCoM), as part of the Computational Materials Sciences Program funded by the U.S. Department of Energy, Office of Science, Basic Energy Sciences, Materials Sciences, and Engineering Division through Argonne National Laboratory. The comparison with experiments and with strained cubic diamond was supported in part by the Q-NEXT Quantum Center, a U.S. Department of Energy, Office of Science, National Quantum Information Science Research Center. Computational resources were provided by the National Energy Research Scientific Computing Center (NERSC), a U.S. Department of Energy Office of Science User Facility operated under Contract No. DE-AC02-05CH11231, and the University of Chicago Research Computing Center.
\end{acknowledgments}

\section*{Data Availability Statement}
The data that support the findings of this study as well as the codes used to analyze it are openly available on \href{https://paperstack.uchicago.edu/}{Qresp}.


\bibliography{lonsdaleite}

\end{document}